\documentclass[11p]{article}
\usepackage{graphicx}
\usepackage{authblk}

\title{Fission of $^{215}$Fr studied with gamma spectroscopic methods}

\author[1]{K. Miernik}
\author[1]{A. Korgul}
\author[1]{W. Poklepa}
\author[2]{J.N. Wilson}
\author[2]{G. Charles}
\author[3]{S. Czajkowski}
\author[1]{P. Czy\.z}
\author[1]{A. Fija{\l}kowska}
\author[4]{L.M. Fraile}
\author[1]{P. Garczy\'nski}
\author[2]{K. Hauschild}
\author[2]{C. Hiver}
\author[3]{T. Kurtukian-Nieto}
\author[2]{M. Lebois}
\author[4]{M. Llanos}
\author[2]{A. Lopez-Martens}
\author[3]{K.M. Deby Treasa}
\author[2]{J. Ljungvall}
\author[2]{I. Matea}
\author[1]{J. Mielczarek}
\author[4]{J. R. Murias}
\author[2]{G. Pasqualato}
\author[1]{A. Skruch}
\author[1]{K. Solak}
\author[2]{K. Stoyachev}
\author[3]{I. Tsekhanovich}
\affil[1]{Faculty of Physics, University of Warsaw, 02-093 Warsaw, Poland}
\affil[2]{Universit\'e Paris-Saclay, CNRS/IN2P3, IJC Laboratory, Orsay, France}
\affil[3]{Univ. Bordeaux, CNRS, LP2I Bordeaux, UMR 5797, F-33170 Gradignan, France}
\affil[4]{Grupo de F\'isica Nuclear \& IPARCOS, Universidad Complutense de Madrid, CEI Moncloa, 28040 Madrid, Spain}

\date{}
\begin{document}

\maketitle

\begin{abstract}
\begin{description}
\item[Background] Asymmetric fission is known to occur in two regions, the
    actinides and sub-lead, and is dependent on the fissioning system excitation
        energy. Experimental evidence in the sub-lead region show that this 
        mode is surprisingly persistent with increasing energy and its origin 
        is not fully understood.
\item[Purpose] To experimentally study the fusion-fission reaction of
    $^{215}$Fr at moderate excitation energy and determine previously 
        unknown independent fission yields and other properties.
\item[Method] The compound nucleus was formed in the reaction $^{18}$O + $^{197}$Au.
    The prompt gamma-rays emitted during the reaction were measured with the high
        efficiency and high granularity $\nu$-ball2 spectrometer. Independent
        fission yields of even-even nuclei were determined by detecting 
        triple-gamma cascades in the fission fragments.
\item[Results] The observed yields, although dominated by a symmetric peak, show
    maxima for heavy fragment of $Z \approx 54-56$, which is consistent with
    the known results in the actinide region but unexpected for the nuclide
        of interest, and at the studied excitation energy.
\item[Conclusions] The mode of asymmetric fission is present even at
    relatively high excitation energies in the system studied. This observation
    matches experimental findings in the sub-lead region, contrary to the 
        actinides,
    and so far there is no well-developed explanation of this phenomenon.
\end{description}
\end{abstract}

\section{Introduction}
One of the commonly observed properties of fission, relevant to the
understanding of this process, is the mass and charge distribution of fragments. During the
process of fission, nuclei can travel along different paths on the potential
energy landscape, which results basically in two general fission modes,
symmetric and asymmetric. The path depends on the fissioning system itself, and
to some degree, on the reaction entrance channel and excitation energy. One of
the key experiments on this matter was performed by K.-H.  Schmidt et al.
\cite{sch00}, which studied 70 systems in the astatine to uranium region via
Coulomb-excitation induced fission.  Symmetric modes were found for systems
with $A < 226$, and asymmetric above that. Another important previous result was the
observation of unexpected highly asymmetric fission of $^{180}$Hg (after
$\beta$-decay) \cite{and10} which triggered a great interest in theoretical
modeling, using various approaches from phenomenological through
microscopic-macroscopic to fully microscopic models
\cite{and13,mcd14,mol15,and16,sch16,sca18,sca19,pom20,mum20}. A remarkable fact
is that most of them could reproduce the observed asymmetric mass split of
$^{180}$Hg, however without common agreement on the physical explanation for it. 

Many models predict asymmetric modes for low excitation energies ($E^* < 20$
MeV), accessible in Coulomb excitation, or $\beta$-delayed induced fission, and
argue that the driving microscopic effects are washed-out at higher excitation
energies \cite{mol15,sch16,sca18,mum20,pom20} resulting in only the symmetric
mode.  However, recent experimental works \cite{nis15,pra20}, show that the
asymmetric mode is clearly seen at moderately high excitation energies in the
sub-lead region. It is quite remarkable that these systems preserve their
asymmetry even for a relatively high excitation energies up to $E^* = 70$ MeV
\cite{nis15}.  Similar phenomena were observed for multi-nucleon transfer
induced fission on U/Th/Pu targets \cite{hir17} up to 60 MeV excitation energy,
but it was explained by a multichance fission mechanism, which has a much
greater impact in the heavier systems. All together there is currently no
unambiguous answer or theoretical consensus on the questions of how the fission proceeds, how it depends
on excitation energy, whether the microscopic structure preserved, and to what
excitation energies. As the same theoretical models or approaches are used for
description of entire chart of nuclides, and applied in astrophysics or
super-heavy elements production planning, it is crucial to stringently test
their predictive powers beyond current experimental knowledge.

In this article we report results of an experimental study of the independent
fission yields, and other characteristics, of $^{215}$Fr, obtained in the $^{18}$O
+ $^{197}$Au reaction. Due to the very short half-live of 101(15) ns \cite{mis19},
this nuclide was not observed in systematic studies of electromagnetic-induced
fission \cite{sch00}, covering francium isotopes with mass number 206--212, and
217--218.  Although the present results, by nature of the reaction used for
the production, probe the fission at much higher excitation energy, this
partially fills in the gap in information about fission in this region.

\section{Experimental setup}
The experiment was performed at the ALTO facitily (IJC Laboratory in Orsay,
France) during the campaign with the $\nu$-ball-2 spectrometer, which is an
upgraded version of the previous $\nu$-ball setup \cite{leb19,wil21,haf21}. The
spectrometer included 24 HPGe Clovers, 10 co-axial HPGe, and 14 LaBr$_3$(Ce)
detectors of (12 cylindrical and 2 conical). The detectors of each type were
arranged in rings - a ring of HPGe detectors, two rings of Clovers, and two
rings of LaBr$_3$(Ce) detectors.  All germanium detectors were equipped with
BGO anti-Compton shields, as well as 5 of the LaBr$_3$(Ce) detectors.  In the
analysis, for various reasons, only 18 clovers and 4 co-axial Ge detectors were
used, along with 12 LaBr$_3$ detectors. The efficiency of the HPGe detectors,
including Compton suppression and addback, reached maximum of a 6\% at 180 keV.
At energies 662 and 1408 keV it was 3.7\% and 2.3\% respectively.  The relative
uncertainty of the efficiency calibration was below 2\% in the range 100--1500
keV, and up to 6\% outside that range, for the energies studied.

A pulsed beam of $^{18}\rm{O}^{7+}$ at the energy 111 MeV, with an average
intensity of 0.5 nA (maximum 2 nA) impinged on a thick, 50 $\mu$m $^{197}$Au
target, with a repetition rate of 400 ns and a pulse width of approximately 2
ns. Data from all of the detectors were collected in triggerless mode by the
FASTER digital electronics acquisition system\cite{FST}, with independent
registration of individual channels based only on a threshold condition. As a
result a total of 16 TB of data were collected during one week of experiment,
and all selection and processing of data was performed off-line.  

The analysis was performed with codes written in the Julia programming language
\cite{bez17} using direct calls to the FASTER digital electronics \cite{FST}
C-library for the raw data format interpretation. The data included the
on-board calculated energy and precise time using the CFD method with the time
bit resolution of 7.8 ps.  The analysis steps included identification of the
RF-signal, the alignement of all channels based on the time differences
compared to the reference detector with the best timing resolution (one of the
conical LaBr$_3$(Ce) detectors), walk correction, and a Compton suppression and
addback procedures applied to the Clover detectors and BGO shielding at common
location. Individual detector hits were then grouped into 400 ns events, in
accordance to the beam pulsation period, but due to the limited time resolution
of the HPGe detectors the event started 50 ns before the beam pulse, and ended
350 ns after. One and two-dimensional spectra such as $\gamma-\gamma$
coincidence, $\gamma$-time or $\gamma$-multiplicity were incremented at that
point, and all events with multiplicity 3 and more (taking into account HPGe
and LaBr$_3$(Ce) detectors) were saved to a pre-processed list-mode HDF5 file
\cite{HDF} including data on energy, time and types of detectors involved. 
This data format allows for the dynamic $\gamma-\gamma-\gamma$ gates that can
use various energy and time conditions depending on the case. In total a
$1.1\times 10^{10}$ threefold coincidences were registered, including
$2.2\times 10^9$ prompt coincidences recorded by the HPGe detectors alone.

\section{General description of the reaction}
The $^{18}$O + $^{197}$Au reaction was previously studied from the point of
view of the production of $^{208-211}$Fr isotopes via fusion-evaporations
\cite{cor05}. In another experiment \cite{app09} the angular distibution of
fission fragments was measured.  Since the excitation functions were well
described by HIVAP code \cite{rei92}, we performed calculations using the same
input parameters as \cite{cor05}.  Our experiment used a thick target, with the
beam completely stopped within. Due to rapidly decreasing fission cross-section
90\% of reactions of that type occurs within first 8 $\mu$m of the target, 
and 99\% within 12 $\mu$m. At the same time, thick target reduces Doppler 
widening of the $\gamma$-rays emitted from the fission fragments.

\begin{figure}
    \centering
\includegraphics[width=1.0\textwidth]{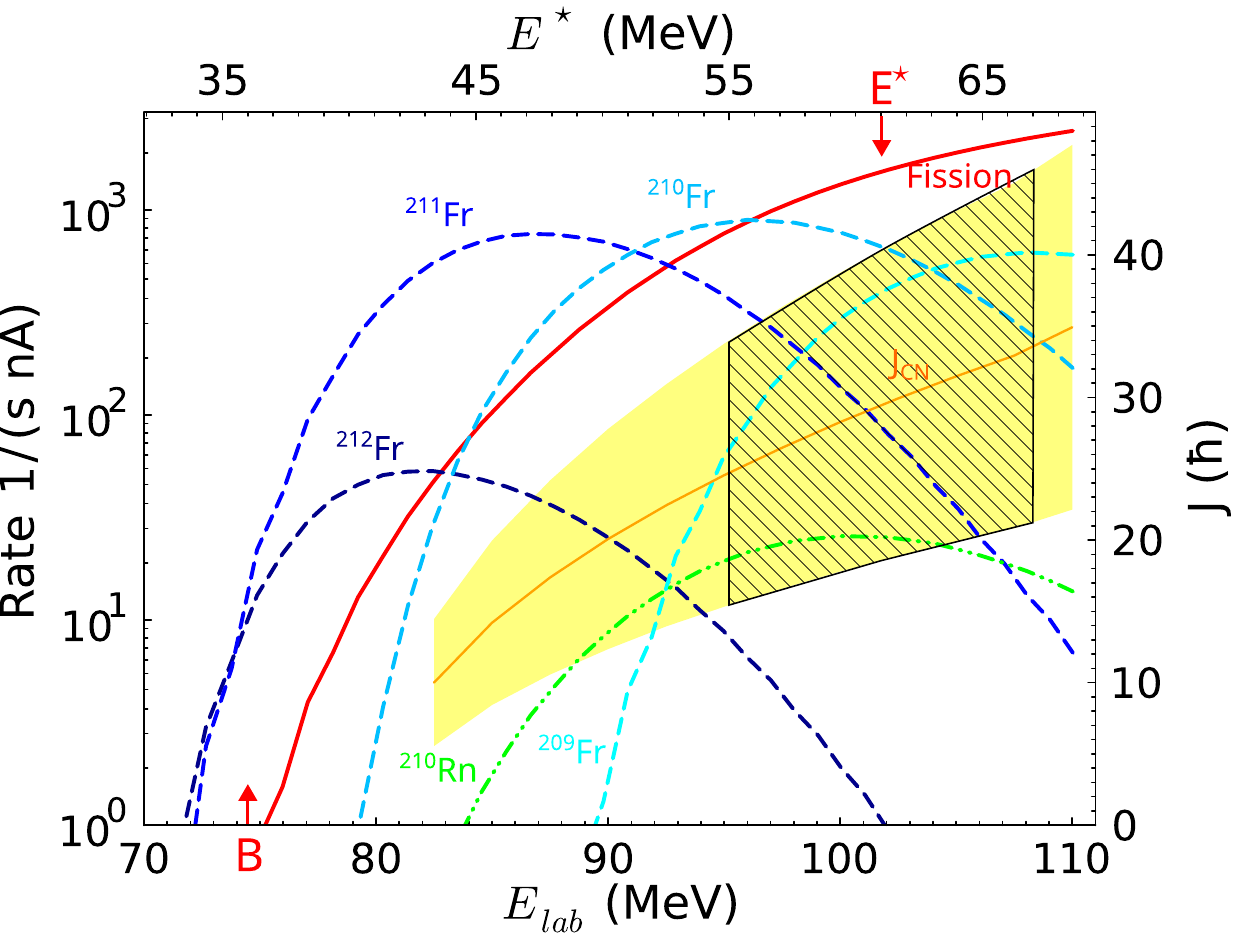}%
    \caption{ Reaction rates per second per 1 nA intensity of
    beam for main reaction channels as calculated by HIVAP \cite{rei92} as a 
    function of the experimental beam energy range (111-0 MeV). Arrows indicate
    effective excitation energy of CN for fission events ($E^\star$) and
    theoretical fusion barrier height ($B$) \cite{mou01}. 
    The orange curve shows the average
    value with the band representing one standard deviation of the total angular
    momentum of the CN as calculated by the PACE4 code\cite{gav80,tar08}.
    The hatched area shows the position of the CN on the $E^* - J$
    space within one standard deviation.
    \label{fig:hivap}}
\end{figure}

To calculate the compound nucleus (CN) excitation energy, we integrated the
expected rates starting from the maximum energy of the beam (111 MeV) down to 0
MeV, including the stopping power \cite{zie10}, and calculating the reaction
rate for each energy step. The weighted mean of energy for each reaction
channel gives the effective excitation energy of the CN, at which the reaction
occurs in the target. The most intense channels are total fission (47\%),
evaporation of 5 neutrons (5n) ($^{210}$Fr 22\%), 4n ($^{211}$Fr 17\%), 6n
($^{209}$Fr 10\%), 3n ($^{212}$Fr 1.0\%), and p4n ($^{210}$Rn 0.7\%).  Together
these channels constitute almost 98\% of all fusion reactions cross-section.
The calculated excitation energy of the CN, at which the fission occurs is
$E^\star = 61.0 \pm 6.2$ MeV with the experimental cross-sections \cite{app09}
and $E^\star = 61.7 \pm 6.1$ MeV with the HIVAP cross-sections. The same
calculations give $49.1 \pm 6.1$ MeV and $56.9 \pm 6.0$ MeV for the 4n and 5n
channels respectively.  The weighted mean rotational energy for the fission
channel was calculated to be $E_{rot} = 4.3\pm2.3$ MeV \cite{sie86}.  Using
PACE4\cite{gav80,tar08} code we have also calculated the angular momentum
distribution of the CN, and the average value, calculated using the same method
as for the excitation energy, was found to be $\left<J\right> = 30 \pm 10\
\hbar$. Due to the sharp decline in the fission cross-section the $E^* - J$
space is constrained to a relatively narrow range, as presented in Fig.
\ref{fig:hivap}, where the hatched area shows this range within one standard
deviation.  For the fission process more relevant is the excitation energy
above the rotating ground state. As the angular momentum decreases with
projectile energy, same is true for the $E_{rot}$, and this effects in even
narrower excitation energy range, with its mean value of $\left<{E^*}'\right> = \left<E^* -
E_{rot}\right> = 57.5 \pm 4.8$ MeV. Theoretical fission barrier for $L = 0$ is
14.8 MeV calculated with macroscopic-microscopic model \cite{mol09}, or 7.8 
MeV with macroscopic model \cite{sie86} using BARFIT subroutine.  The latter
model allows to include the effects of angular momentum of CN, and weighted
barrier height, using the cross-sections and angular
momentum distribution, was found to be lowered to 5.9 MeV.

\begin{figure}
    \centering
\includegraphics[width=1.0\textwidth]{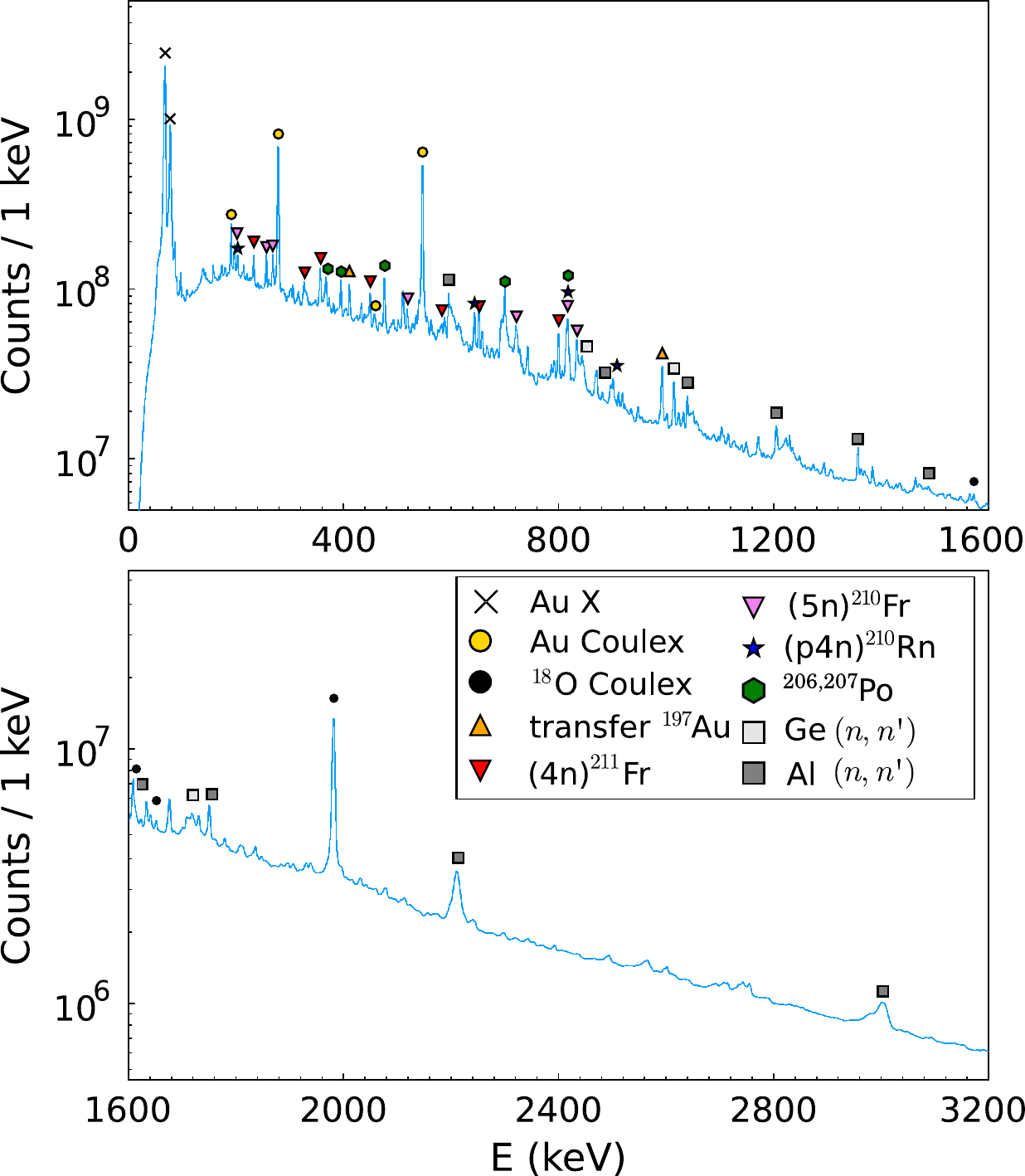}%
    \caption{The total gamma ray spectrum from HPGe detectors obtained
    during the experiment. Some major gamma-lines are marked with
    their origin. See text for more details.\label{fig:gamma}}
\end{figure}

Figure \ref{fig:gamma} presents the total gamma spectrum obtained during the
experiment. The main contributors to the spectrum are Coulomb excitation of the
target ($^{197}$Au) and projectile ($^{18}$O), transfer reactions to and from
the target ($^{196}$Pt, $^{198, 200}$Hg), inelastic neutrons scattering on the
detectors and materials (mainly Ge and Al), fusion-evaporation residues
($^{210,211}$Fr, $^{210}$Rn) and their daughter activities ($^{206,207}$Po).
These lines constitute background for the fission reaction. Most intense lines,
that have potential for the random coincidences, are originating from the
Coulomb excitation reaction, but their impact can be reduced by requiring a
higher multiplicity ($\geq 3$ HPGe). Transfer and fusion-evaporation reaction
can populate high-spin states and result in the high $\gamma$-multiplicities,
but are less intense, and there is a marginal probability of their random
presence in a threefold coincidences. Most care needs to be taken with the $(n,
n')$ reactions, which generate true coincidences, as the neutrons are indeed
emitted during the fission. However, their characteristic shape and known energies
\cite{bag20}, allow their recognition in the fission fragment gamma ray spectra.

\section{Analysis}
The traditional experimental methods to determine fission yields make use of
direct identification of the fragments mass or charge via the TOF/$\Delta$E
detectors or spectrometers \cite{sch16}. An alternative method, based on the
detection of the prompt $\gamma$ radiation from fragments, known already in the
70's \cite{che71} has been recently developed for use with large spectrometers
giving promising results \cite{bog07,wil17,wil18,mus20,dey21,leo22}. This
method, used during the first $\nu$-ball campaign at ALTO, was proven very
successful at identifying fragments, and giving insight into angular momentum
generation \cite{wil21}. It provides great selectivity, as there is unambiguous
fragment identification, and also gives the possibility to extract other
information such as gamma multiplicity, angular correlations, or total gamma
energy, which are also of interest.  The biggest drawback is, however, its
inability to detect small de-excitation branches and direct feeding to the
ground state or the first excited state. These factors can potentially
introduce systematical errors, and limit the accuracy of the method. 

The literature shows a number of examples of fission fragments yields being
determined based on the gamma spectroscopy. These are almost exclusively
limited to the even-even isotopes, with well known low-lying level
structure. Methods are based on the detection of the gamma-rays from the first
$2^+$ state ($2_1^+ \rightarrow 0^+_{gs}$) \cite{che71}, coincidence of this
transition with the $4^+_1 \rightarrow 2^+_1$ \cite{dey21} or all possible
transitions feeding the $2^+_1$ state \cite{ban15,wil17}. Other approaches rely on
coincidence of the $2_1^+ \rightarrow 0^+_{gs}$ or $4^+_1 \rightarrow 2^+_1
\rightarrow 0_{gs}^+$ with the $\gamma$-rays emitted by even-even fission fragment
partners (i.e. $2_1^+ \rightarrow 0^+_{gs}$ or $4^+_1 \rightarrow 2^+_1
\rightarrow 0_{gs}^+$) \cite{bog07,mus20,dey21}, or detection of the $6^+
\rightarrow 4^+_1 \rightarrow 2^+_1 \rightarrow 0_{gs}^+$ cascade \cite{dey21}.
The systematical errors of these methods depend on the type of reaction (e.g.
population of the excited levels is different in the spontaneous fission,
neutron induced fission and heavy-ion induced fission), chosen transitions, and
quality of the nuclear structure data, including fragments level
schemes, level half-lives etc. Depending on the statistics one can expect the
systematical uncertainties to be on the level of 10--30\%
\cite{bog07,wil17,dey21}, typically comparable with the uncertainties
associated with the efficiency calibration and statistical errors \cite{wil17}.

\begin{figure}
    \centering
\includegraphics[width=0.7\textwidth]{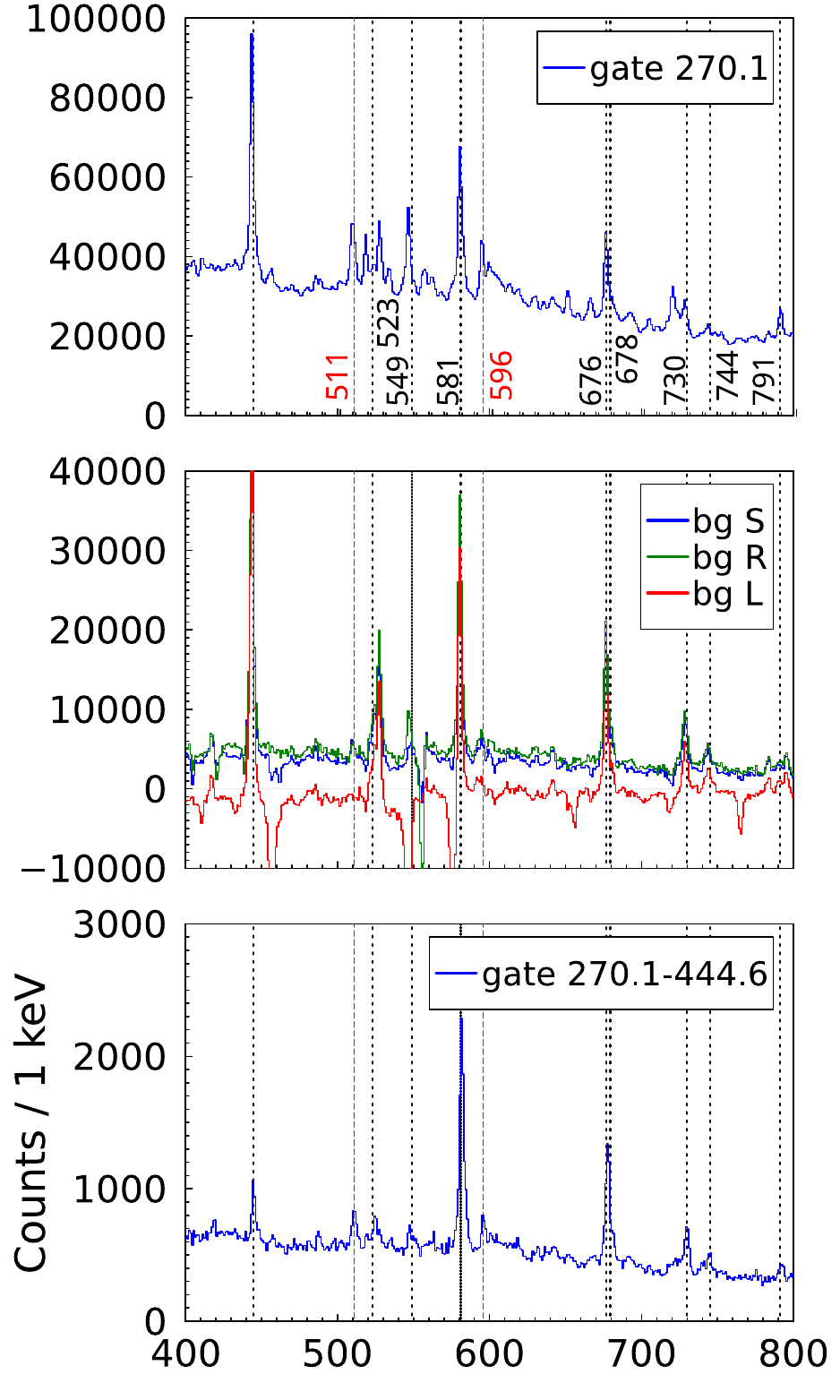}%
    \caption{
    Example of $\gamma$-$\gamma$ spectrum (top), $\gamma-\gamma$ spectrum with
    background subtraction (middle) and $\gamma-\gamma-\gamma$ spectrum
    (bottom) gated on the known 270 keV ($2^+\rightarrow 0^+$) or 445 keV ($4^+
    \rightarrow 2^+$) transitions in the $^{106}$Ru.  Black dashed lines show
    position of the identified transitions, red dashed lines show the neutron
    scattering lines and the $e^+e^-$ annihilation line.  See text for more 
    details.
    \label{fig:gates}}
\end{figure}

The initial system in our study ($^{215}$Fr) is an odd-Z nuclide, therefore
methods based on the correlations between fission fragment gamma rays could not be
applied.  Considerable background coming from reactions other other than fission
(c.f Fig. \ref{fig:gamma}) imposed the need to use  more selective methods than
single or twofold coincidences. Our method is therefore based on threefold
$\gamma-\gamma-\gamma$ coincidences using all known transitions leading to the
ground state of a given fragment. The selectivity of this method is shown in
Fig. \ref{fig:gates}. Gamma-gamma coincidence spectra (Fig. \ref{fig:gates}
top) is not sufficient to distinguish between transitions in the $^{106}$Ru and
in other isotopes (specifically $^{210}$Fr which has a gamma line at a similar
270 keV energy).  Methods based on the $\gamma-\gamma$ coincidences with
background subtraction, may suffer from the ambiguities arising from the way
the background region is selected. The middle panel of the Fig. \ref{fig:gates}
shows three methods of background subtraction using the gate of the same width
as the line gate. These are: symmetric on the both sides of the peak (blue),
located to the right (green), and to the left of the peak (red). The intensity
or existence of some of the lines may depend on the chosen method. In the
$\gamma-\gamma-\gamma$ gate, (Fig. \ref{fig:gates} bottom) shown without
background subtraction, only the known lines attributed to the $^{106}$Ru along
with known $\gamma$-rays originating from the inelastic neutron scattering on
the detectors and surrounding material and the $e^+-e^-$ annihilation line are
visible. 

The detailed description of the procedure for extraction of the fission
fragments yields is given below.  We first selected a set of even-even nuclei
in the expected region of reaction residuals i.e. $Z = 26-64$ for fission, and
$Z = 78-86$ for transfer, and fusion-evaporation reactions, with $N/Z$ ratio
between 1.2 and 1.6 ($N/Z$ for $^{215}$Fr is 1.47). For all these nuclei, all
cascades in the ENSDF database \cite{NNDC}, of three $\gamma$-rays were
selected, under the condition that the transitions are directly connected, and
that the last transition is feeding the ground state. For each unique cascade,
 $\gamma-\gamma-\gamma$ threefold coincidence spectra were extracted from the
experimental data, by setting a condition on the two bottom transitions and
requiring that the $\gamma$-rays were emitted within the prompt gate (i.e.
$\pm$15 ns, relative to the beam pulse). Each spectrum was then inspected to
search for the third, feeding, transition.

\begin{figure}
    \centering
    \includegraphics[width=0.65\textwidth]{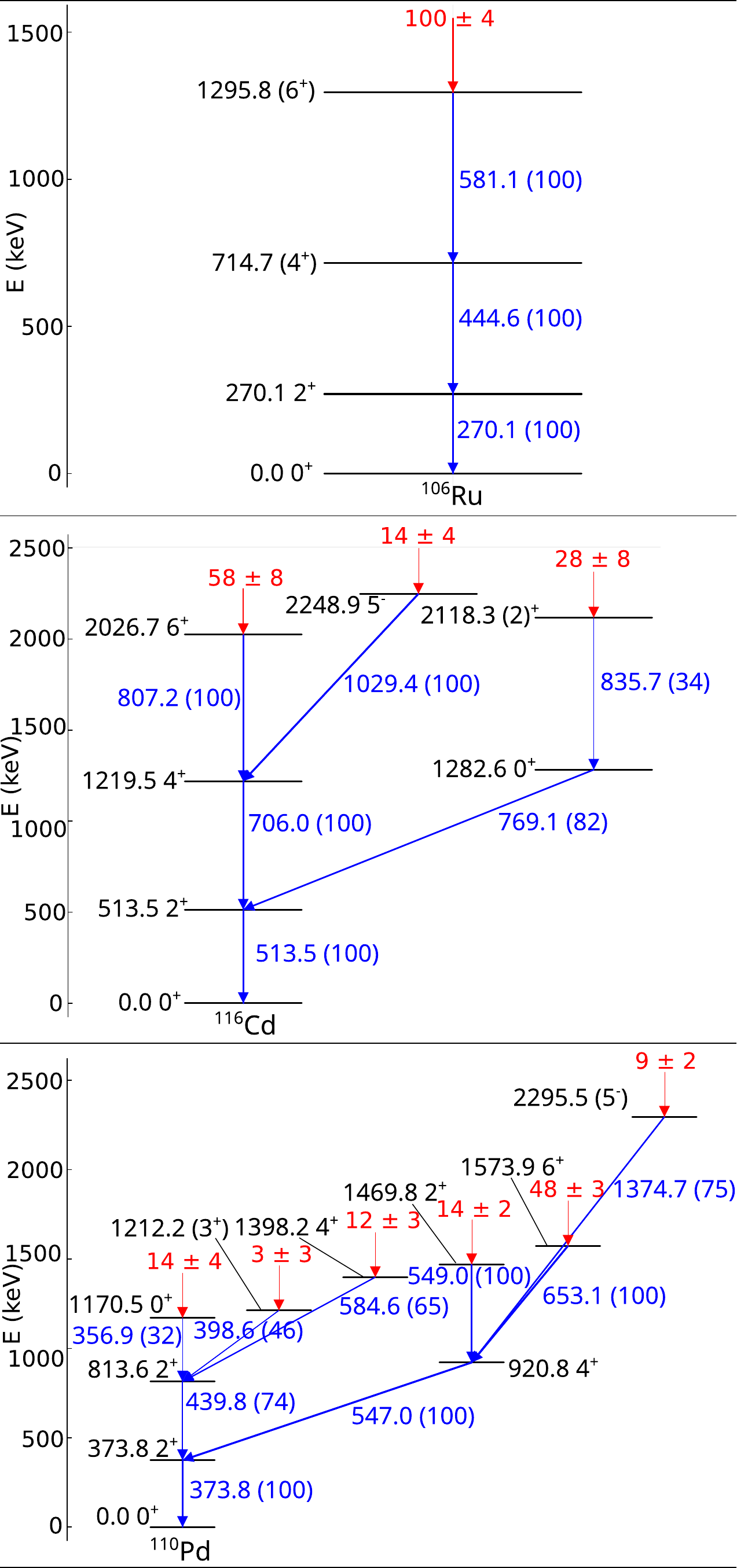}%
    \caption{Example of the observed feeding patterns of fission
    fragments.  Black numbers indicate levels and their spin. Blue values show
    gamma transitions and their relative intensity in brackets. 
    Red labels and arrow show observed feeding to the indicated state.
    \label{fig:patt}}
\end{figure}

As an example, figure \ref{fig:patt} presents results obtained for three
nuclides. In $^{106}$Ru only the $6^+_1 \rightarrow 4^+_1 \rightarrow 2^+_1
\rightarrow 0^+_1$ cascade is observed. The existence of low-lying $0^+_2$ in
$^{116}$Cd introduces additional fragmentation apart from commonly observed in
other nuclei feeding from $5^-_1 \rightarrow 4^+_1 \rightarrow \ldots$.
Finally, $^{110}$Pd shows greatest fragmentation of the observed feeding, where
it is spread over 6 levels, and even though the $6^+_1$, as expected,
is dominant, it yields only 48\% of the feeding.

A total of 147 isotopes, 2561 gates, and 13591 lines were analyzed with an
automatic fitting procedure. The conditions for accepting a line were: energy
within $\pm$~1~keV compared to the value known from database, correct line
width ($\sigma$ = 0.5 - 1.5 keV, energy dependent), area of the peak larger
than 0 within the 3$\sigma$ test, and negative result of the Kolmogorov-Smirnov
statistical test at 5\% confidence level. The latter condition was testing the
hypothesis that the distribution of counts within the expected peak area is
from the same distribution as the counts area outside of the peak. The results
of the automatic procedure were next critically inspected, and in some cases
corrected, mostly due to a doublets, nearby line, or a $\gamma$-ray originating
from the $(n, n')$ reaction with energy close to the expected value for the
analyzed line.

The intensities obtained were corrected for the half-life of the following
isomeric states in the range of 1~ns -- 1~$\mu$s: $^{92}$Zr ($6^+/5^-$), $^{106}$Pd
($5^+$), $^{120,122}$Sn ($5^-$), $^{130}$Xe ($6^+$), and $^{138}$Ba ($6^+$).
Three longer lived isomeric states above the analyzed cascades were also
identified: $^{84}$Kr ($8^+$) 1.83~$\mu$s, $^{120,122}$Sn ($7^-$) 11.8~$\mu$s
and 7.5~$\mu$s. In these cases, the intensities were corrected based on the
time gate opened for the whole cycle. Next, we obtained corrections for
the $\beta$-decay contribution for all the studied isotopes using the
coincidences in the delayed window in respect to the beam pulse. The only
ambiguity has arisen in the case of $^{120,122}$Sn, were the isomeric $7^-$
states are also populated in the $\beta$-decay of $8^-$ isomers of
$^{120,122}$In. As the ratio of the population of isomers to the ground states
in fission is not known, and the states half-lives are substantially longer than
the beam pulsing, it is not possible to determine what part of the intensity
should be attributed to the $\beta$-decay. Since the population of the
high-spin isomers is expected to be smaller than the ground state, in these
cases we assumed that the contribution from $\beta$-decay is negligible, but an
additional systematical error may have been introduced here.

Using the information on the transitions absolute intensities we were able to
calculate the absolute population of the levels of interest in the fission
fragments, and following that, the total fragment population. If cascades were
originating from the same initial level, the feeding was calculated as the
weighted average of all the cascades starting from the level. In total we found
a 144 level feedings in 69 isotopes, including 60 fission fragments. See
Supplemental Material at [URL] for a detailed table with numerical values for
observed number of individual fission fragments. The results are discussed in
the detail in the next section. 

One of the main limitations of the spectroscopic method of yield determination
lies in the inability to detect the ground state feeding. Using the threefold
coincidences based on the transitions in only one of the fission partners, we
are also unable to detect the direct feeding to the first and second excited
states in the cascade. As it was observed in \cite{dey21}, the population above
$4^+_1$ level tends to be fragmented, and relaying solely on $6^+_1 \rightarrow
4^+_1 \rightarrow 2^+_1 \rightarrow 0^+_1$ is therefore unreliable. The method
presented here overcomes this limitation by analyzing all kinds of transitions,
and possible systematic errors are coming only from the direct population of
the lowest levels, with events of multiplicity of 2 and less.  While this side
feeding can be important \cite{wil21}, it results in the systematical errors
only if there are significant differences between isotopes. Such effects were
observed in the neutron induced fission of $^{238}$U \cite{wil17}, and to some
extent in the heavy-ion induced fission ($^{18}$O + $^{208}$Pb) \cite{bog07}.

In order to determine the impact of this effect we used a similar method as
described before, but based on the $\gamma-\gamma$ coincidences for all
possible known transitions de-exciting to the ground state. This method is
therefore able to detect the direct feeding to the second excited state in a
given cascade, and possible differences between the two methods may indicate
the aforementioned effect. Unfortunately gamma-gamma analysis suffers from a
worse selectivity then that based on threefolded coincidences, and a lower
sensitivity to weak transitions that may be ambiguous to identify. For the
fragments clearly identified by the both methods we calculated the ratio of the
yields obtained ($r = Y_{\gamma\gamma}/Y_{\gamma\gamma\gamma}$). A statistical
t-test failed to reject the hypothesis that this ratio is uncorrelated with the
mass or charge of the fragments. The mean value of $r$ was found to be
1.0$\pm$0.35, while the mean relative statistical uncertainty for both methods
was at the level of 25\%. As a result we conclude that the possible
systematical uncertainty due to the undetected direct feeding to the low lying
states is lower than the statistical errors and should not affect the
distributions obtained in a considerable way.

\begin{figure}
    \centering
    \includegraphics[width=1.0\textwidth]{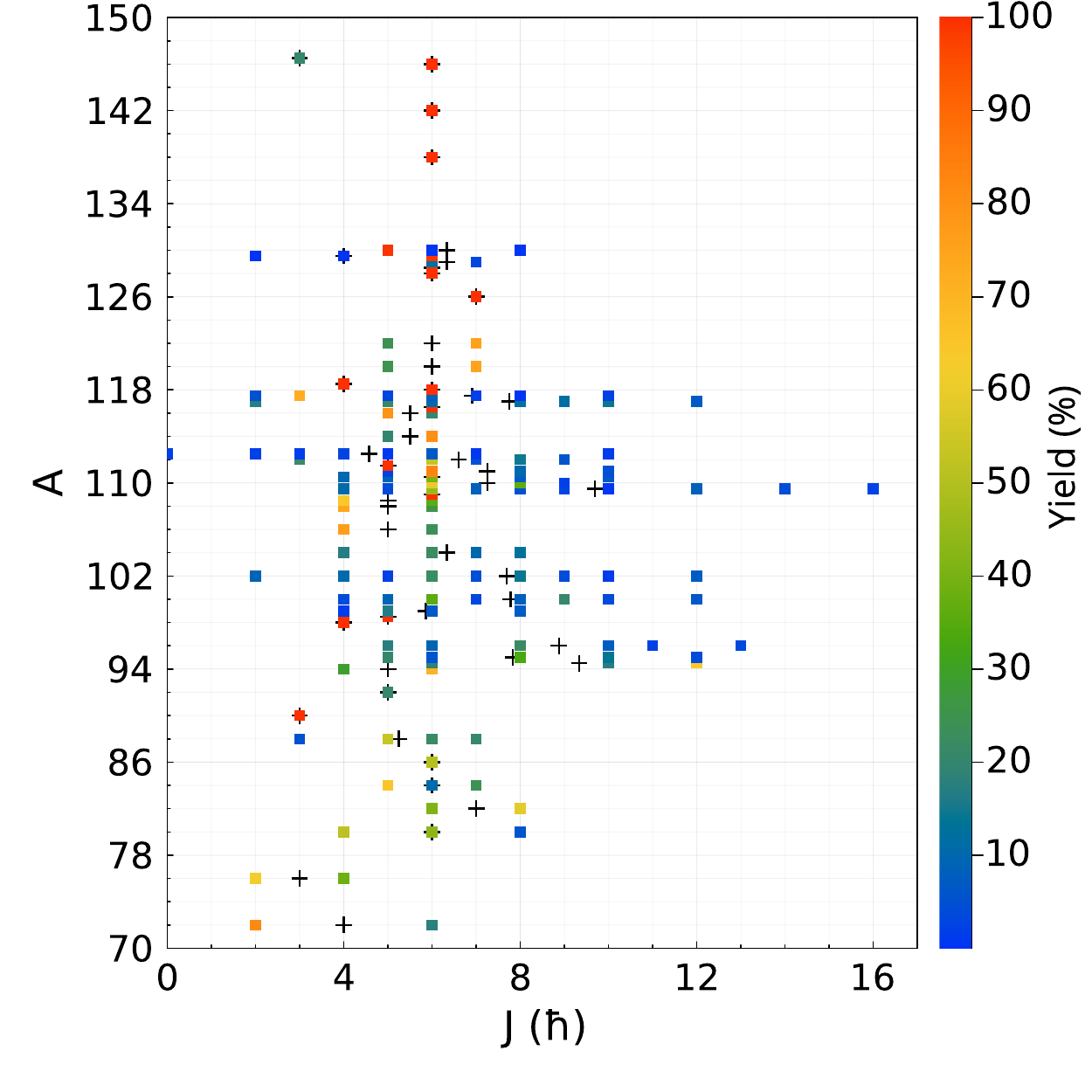}%
    \caption{
    Individual fragments levels population in function of mass. Crosses show
    calculated average spin per fragment. Fragments of the same mass are
    sorted by their charge, and shown next to each other. See text for 
    more details.
    \label{fig:spin}}
\end{figure}

By applying the Manchester method \cite{abd87}, we have also calculated the
average spins of the individual fragments, by detecting all possible
transitions above the states used in the fragment yields determination. In this
case the analysis is limited by the available knowledge of the excited states
in the fragments.  In a number of cases we have found transitions from states
with an unknown spin assignment, or with a suggested range or lower limit only.
Those states were excluded from the analysis, but it clearly affects the
overall picture, due to systematical effects connected with more complete level
schemes of some the fragments e.g. those studied in fission of $^{252}$Cf or
$^{235}$U.  The results presented in Fig.\ref{fig:spin} show the discrete
levels population in individual fragment in the function of their mass. The
yield is normalized to 100\% for each fragment.  For the masses with more than
one fragment detected, the results are sorted by the fragment charge. The black
crosses show the calculated average spin of each fragment. The overall mean
spin per fragment was found to be $\bar{J} = 5.9\pm1.5\ \hbar$. As a
consequence of the aforementioned issues, this result should be treated as a
general guidance and qualitative description more than the quantitative value.
Our results are close to those described in \cite{abd87}, where a similar
reaction was studied ($^{19}$F + $^{197}$Au).

\section{Discussion}

In order to compare results with theoretical calculations and other
experimental results, the yield obtained were normalized so that the sum of
observed fragments is equal to 50\%. This assumes that there are no
systematical differences between even-even, even-odd and odd fragments, and the
impact of non-observed even-even fragments is negligible. Minimal and maximal
observed yields for individual fragments were 0.01(3)\% and 3.4(2)\% for
$^{118}$Sn and $^{102}$Mo respectively.

\begin{figure}[h!]
    \centering
\includegraphics[width=1.0\textwidth]{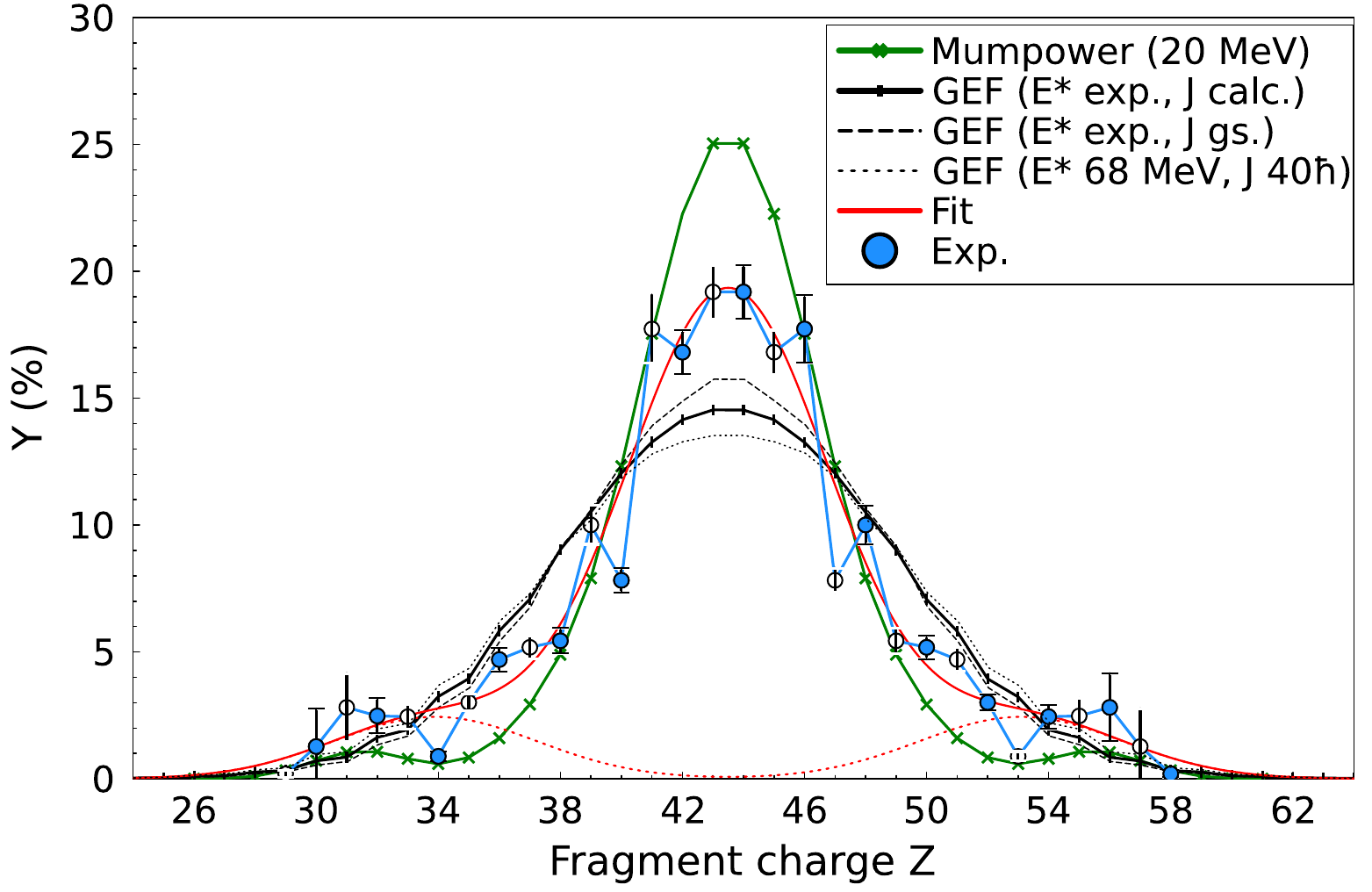}%
    \caption{Experimental and calculated fragment charge
    distributions (see text for more details). \label{fig:charge}}
\end{figure}

Fragment yields arising from the fission of isotopes in the vicinity of
$^{215}$Fr were studied previously by measuring the fragments charge
\cite{sch00}, these included $^{206-212}$Fr, $^{217-218}$Fr, $^{209-219}$Ra.
Fission fragment mass distribution from nearby $^{213}$At \cite{itk90} and
$^{214}$At \cite{pau20} were also studied. The francium isotopes were found to
breakup in a symmetric way, but $^{214-219}$Ra have a measurable asymmetric
component, largest for $^{216}$Ra, lying just one proton above $^{215}$Fr. The
mean position of the heavy fragment in asymmetric fission was shown to be $Z
\approx 54$ all across the measured range. Figure \ref{fig:charge} shows the
charge distribution obtained in our experiment. As the number of protons in the
fission is preserved, the experimental points were symmetrized, and yields for
the fragments with complementary charge $Z_s = 87 - Z_i$, are shown with the
open symbols. This method not only fills in the missing data points for the
odd-$Z$, but presents the self-consistency of the data. Systematical issues
with the yield determination method might potentially break the symmetry, which
is not observed here.

It is clear that the experimental data cannot be described by a single
Gaussian distribution, as on the both sides of the main maximum, there are
additional tails, around $Z \approx 54-56$, and $Z \approx 30-32$. Following
the procedure described in \cite{sch00} we fitted a triple Gaussian shape to
the experimental distribution using both the unconstrained fit, and a fit with
the fixed position for heavy fragment at the $Z_h = 54$ and light at the $Z_l =
33$. Both results are consistent within the uncertainties. The unconstrained
fit gives the position of the asymmetric peaks at $Z_h = 53.1\pm1.9$ and $Z_l =
33.9 \pm 1.9$. The ratio of symmetric to asymmetric component was found
$Y_s/Y_a = 7.6 \pm 1.9$ ($6.8 \pm 1.1$ with the constrained fit).  Results of
the unconstrained fit are shown in Fig. \ref{fig:charge}.  While these results
cannot be directly compared to the systematics in \cite{sch00}, because of a
different compound excitation energy and angular momentum, the $Z \approx 54$
asymmetric component is still present, even at such a high excitation energy of
$\left<{E^*}'\right>= 57.5 \pm 4.8$ MeV, well above the fission barrier. 

Figure \ref{fig:charge} presents results obtained with the GEF code
\cite{sch16}. The fission fragments yield was calculated using distribution of
energy and angular momentum states of the CN. For each energy step, a FF
distribution was calculated for a predicted range of $J$ values. The presented
result is a weighted average over excitation energy and $J$ matching the
experimental distribution, and PACE4\cite{gav80,tar08} calculations.
Distributions calculated for the $L = 0$ (i.e. total angular momentum equal to
the ground state), and for $L = 40\ \hbar$ and $E^* = 68$ MeV are also shown
for a reference.  The GEF code takes into account angular momentum by modifying
fission decay-width (\cite{sch16}, Eqs. 85-87), as well as symmetric
fission channel width which is calculated based on a modified empirical
systematics (\cite{sch16} Eq.  61).

For comparison a results of microscopic-macroscopic model by M. Mumpower et al.
\cite{mum20}, for which the distribution is available only at an excitation
energy slightly above the fission barrier, and ground state $J$, are also
presented. The latter model does not include post-fission de-excitation by the
neutron emission, but the charge distribution should be preserved. Both models
give the individual isotopic yield, so the experimental data normalization can
be verified. By selecting even-even isotopes which are predicted to have yield
larger than the experimentally observed minimal value, we obtain the sum of
0.499 for both the GEF and Mumpower models. Due to different excitation
energies, the quantitative comparisons are limited, nevertheless the
experimental distribution is similar to those predicted by both models. While
the GEF predicts a single, wide Gaussian shape (growing wider with an
additional angular momentum and energy in the system), the M. Mumpower model
shows a narrower central distribution with the asymmetric components in similar
locations to the experimental values, however, with a smaller amplitude.  

It is worth noticing that other microscopic models, e.g. \cite{mol15,ghy14},
also show significant asymmetric component in mass distributions for the low
energy fission, but lack of charge and full fragments distribution information
precludes making detailed comparisons with our data. Nevertheless this fact may
indicate that combination of microscopic model, with proper treatment of
excitation energy and angular momentum could be sufficient to explain observed
distributions.  This should include the interplay of increasing probability of
energy and angular momentum removal by pre-scission neutrons with excitation
energy and J, and a increasing fission barrier height with decreasing spin and
possible appearance of asymmetric mode at high $E*$ and $J$ states.

\begin{figure}[!ht]
    \centering
\includegraphics[width=1.0\textwidth]{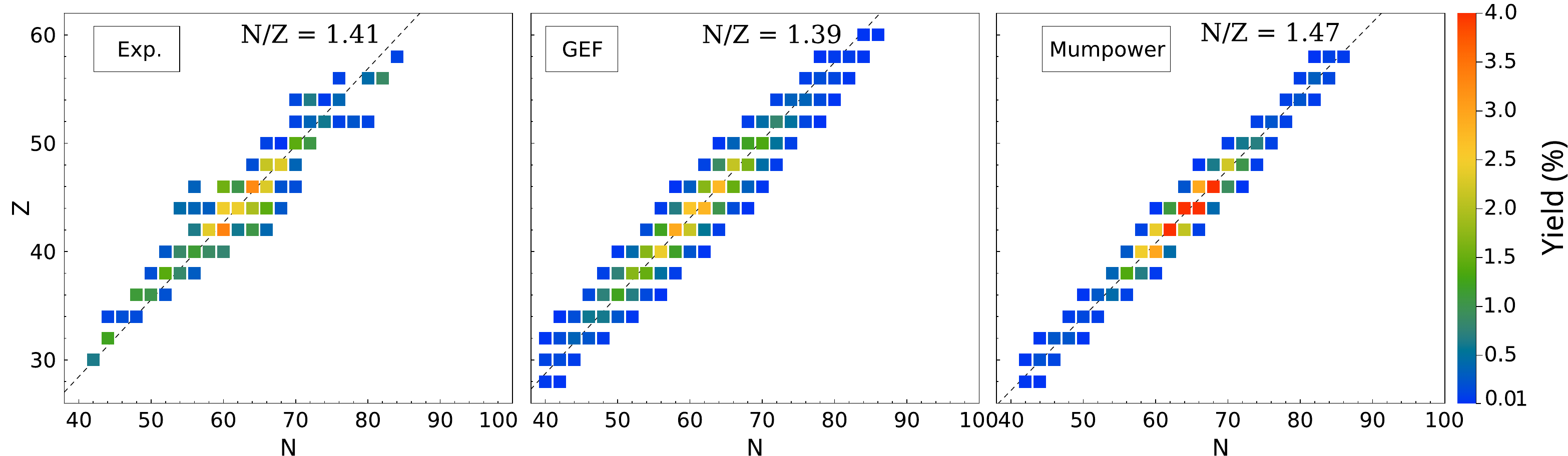}%
    \caption{
    Left: experimental fragment yield, color encoded.  Middle: fragment yield
    calculated by GEF model \cite{sch16} for experimental excitation energy
    and angular momentum. Right: fragment yield calculated by Mumpower et
    al. \cite{mum20} for excitation energy 20 MeV and $L = 0$.  Dashed lines
    show calculated ratio of $N/Z$ for the given distribution. See text for
    more details.  \label{fig:maps}} 
\end{figure}

The method based on gamma spectroscopy offers possibility to inspect fragment
yields, without losing information due to projections on $A$ or $Z$ directions.
This is shown in Fig. \ref{fig:maps}, where the experimental distribution and
those calculated with GEF and by M.  Mumpower models are presented. For clarity
the results of the models show only the even-even isotopes with yield above the
experimental limit. The dashed lines show the calculated $N/Z$ ratio weighted
by the individual yield of the fragments.

The asymmetric component is again seen in the experimental results for $Z =
52-56$, and $Z = 32,34$. These proton numbers are known to be located around
deformed closed shells \cite{naz85,lea85,sch00,sca18,sca19}. The mean $N/Z$
ratio for the experimentally observed fragments is 1.41, which indicates on
average emission of 5 neutrons per fission.  The results of M. Mumpower are
more confined along $N/Z$ line and shifted to the more neutron-rich side ($N/Z
= 1.47$, equal to CN), which is perfectly understandable, as this model
predicts only primary fragments, before the neutron emission. The GEF model
distribution, which include the neutron emission from the CN and the fragments,
is shifted to the left, with $N/Z = 1.39$, which translates to an average of
7 neutrons emitted per fission.

The direct comparison of the experimental results with the theoretical models
is even more difficult due to the role of the multichance fission. The GEF code
predicts a significant contribution of this mode, as only 24\% of events are
the first-chance fission, and following chances are 14\%, 27\%, 16\%, 15\%, and
3\% for up to 5 pre-saddle neutrons emission respectively, with an average of 2
pre-fission neutrons emitted.  Multi-chance fission probability is decreasing
with increasing angular momentum, but the overall impact is minimal. According
to the GEF calculations for the excitation energy $E^* = 66$ MeV,
the first-chance fission for $J = 3/2 \hbar$ is 17.7\%, while for $J = 79/2$ is
18.4\%. For the excitation energy $E^* = 52$ MeV, these values are 41.5\% and
40.4\% respectively. The details of distribution of excitation energy above
rotating ground state, CN angular momentum, and fission barrier height, 
calculated with GEF\cite{sch16}, PACE4\cite{gav80,tar08}, and BARFIT\cite{sie86}
are given in Table \ref{tab:multi}.
\begin{table}
    \centering
\caption{Average values and standard deviations of excitation energy 
    $\left<{E^*}'\right>$, CN angular momentum $\left<L\right>$, and fission
    barrier $\left<B_f\right>$ for multi-chance fission channels. \label{tab:multi}}
\begin{tabular}{ccccc}
    \hline\hline\\
Chance & Prob. (\%) & $\left<E^{*'}\right>$ & $\left<L\right>$ & $\left<B_f\right>$ \\ \hline
  1 &  24.2 &   $56.2 \pm 5.0$ &  $28.7\pm 10.0$  &   5.8\\
  2 &  14.3 &   $56.6 \pm 4.9$ &  $29.0\pm 10.0$  &   5.6 \\
  3 &  27.0 &   $57.2 \pm 4.8$ &  $29.3\pm  9.9$  &   5.5 \\
  4 &  16.5 &   $58.2 \pm 4.5$ &  $29.9\pm  9.9$  &   5.2 \\
  5 &  14.8 &   $59.8 \pm 3.5$ &  $30.7\pm  9.7$  &   5.0 \\
  6 &  3.0  &   $60.7 \pm 2.9$ &  $31.0\pm  9.7$  &   4.8 \\
    \hline\hline\\
\end{tabular}
\end{table}

The multichance fission mechanism would however inevitably drive the system
towards the nuclei which are known to follow the symmetric mode even at low
excitation energies, such as $^{212-210}$Fr \cite{sch00}, which cannot be the
source of the asymmetric tails. As a result, isotopes $^{213-215}$Fr must have
a considerable asymmetric mode contribution even at moderate excitation
energies.  The multichance fission along with impact of the angular momentum
cannot be the only explanation of the observed asymmetric fission mode, as in
the case of isotopes of uranium \cite{hir17}.  One must also remember that the
GEF does not predict a significant role for the asymmetric mode, and
overestimates the number of emitted neutrons, which may indicate that the
multichance fission impact is lower than that calculated by this model, and
origin of the experimentally observed behavior is due to local microscopic
effects, beyond statistical and phenomenological approach of the GEF code.

Most of the microscopic model calculations are available at low excitation
energy only \cite{ghy14,mol15,mum20}, and in all cases the authors assume that
the asymmetric modes, appearing due to shell effects, should quickly disappear
with excitation energy. This statement is common in other approaches e.g.
scission point models \cite{car19,pas23} or microscopic energy density
functional model \cite{mcd14,sca18}.  While the disappearance of the asymmetric
mode is confirmed by the experimental studies in the heavier nuclei e.g.
$^{226}$Th \cite{rus08} or even slightly lighter e.g.  $^{214}$At \cite{pau20}
our observation is on par with observations in the lighter systems, such as
isotopes in the sub-lead region \cite{nis15,tse19,pra20,bog21,koz22,kum23},
where this mode was shown to be persistent. This might indicate that the
disappearance of the asymmetric mode is not universal and neither constrained
to the selected regions of the chart of nuclides.  This fact is supporting the
general connection of the asymmetric mode and deformed shells, and a common
mechanism of this mode for both the actinides and the sub-lead regions
\cite{ich19,sca19}.  

\section{Summary}
Previously unknown properties of the fissioning system $^{215}$Fr at an
excitation energy of 61.0 MeV were studied. It was formed as a compound nucleus
in $^{18}$O and $^{197}$Au reaction. Methods based on the spectroscopy of the
prompt gamma radiation was used to identify the even-even fission fragments.
The obtained results may serve as input for planning of the production and
spectroscopic studies of a moderately exotic nuclei with heavy-ion induced
fission.  We have found a significant asymmetric component in the charge
distribution, yielding $11.6 \pm 2.9$ \%, and located for the heavy fragments
around $Z \approx 54-56$. This finding is consistent with a known behavior of
nuclei in the actinide region, but rather unexpected at the present excitation
energies.  The multichance fission mechanism cannot be the only explanation to
this phenomena, and $^{215}$Fr and neighboring nuclei must have a significant
asymmetric mode at moderate excitation energies. There are some conclusions
from the microscopic models on the driving force behind asymmetric fission, but
dependency on the excitation energy and angular momentum is not fully studied
or understood. Our results may call for developments in the description of the
fission process as well as new experimental studies which could explore
different excitation energies or using neighboring francium isotopes as the
fissioning system.

\section{Acknowledgments}
This work was partially funded by the Polish National Science Center under Grant No.  2020/39/B/ST2/02346. 

This work has been funded by by the European Union's Horizon 2020 research and innovation programme Grant Agreement No. 654002 (ENSAR2). Funding by Spanish AEI via grants RTI2018-098868-B-I00 and PID2021-126998OB-I00 is acknowledged.

\bibliographystyle{ieeetr}
\bibliography{refs.bib}

\end{document}